\documentclass[12pt,onecolumn,showpacs,amssymb,aps,nofootinbib,floatfix]{revtex4-1}

\usepackage{epsfig}
\newcommand{\ave}[1]{\left\langle #1 \right\rangle}

\newcommand{\order}[1]{ \mathcal{O} \left( #1 \right) }
\newcommand{\eqcomma}{\phantom{AA},\phantom{AA}}
  \newcommand{\lqcd}{\Lambda_{\mathrm{QCD}}}

\begin{document}

\title{On the origin of long-range azimuthal correlations in hadronic collisions}


\author{Giorgio Torrieri$^{a,b,c}$}
\affiliation{$\phantom{A}^a$Instituto de Fisica Gleb Wataghin, Universidade Estadual de Campinas  UNICAMP, Campinas, SP, 13083-859 Brazil\\ $\phantom{A}^b$FIAS,  Johann Wolfgang Goethe-University, Frankfurt am Main, 60438, Germany\\ $\phantom{A}^c$Pupin physics laboratory, Columbia University, New York, 10027,USA}

\begin{abstract}
I review the models suggested, to date, as an explanation for the so called ``ridge'' phenomenon, an elongation in rapidity of 2-particle correlations seen at RHIC and LHC energies. I argue that these models can be divided into two phenomenologically distinct classes:  ``Hotspot+flow'' driven correlations, where initial state correlations created by structures local in configuration space are collimated by transverse flow, and models where the azimuthal correlation is created through local partonic interactions in a high gluon density initial state.

I argue that the measurement of a strong double ridge in $pA$ and $dA$ collisions allows a good opportunity to understand the ridge's origin because it allows to  see if a common Knudsen number scaling, expected if the ridge has a hydrodynamic origin, can be used to understand all data.
I show that current data presents evidence that this scaling is lacking, presenting a challenge to the hydrodynamic models. 

On the other hand, particle-identified correlations are a particularly promising way of testing the assumption, which distinguishes the two models, of whether the correlation is formed initially in the partonic phase, or as a final state effect.
Assuming fragmentation occurs ``as in vacuum'' can be used to predict scaling trends which are generally broken by models, such as hydrodynamics, where the ridge is created as a final state effect.  While evidence is again not fully conclusive, data does seem to follow a scaling compatible with hydrodynamics \cite{us}.

I close by discussing experimental observables capable of clarifying the situation
\end{abstract}

\maketitle
\section{Introduction}
The so-called ``ridge'', found in hadronic collisions at RHIC and LHC energies, 
\cite{brodsky,aaridgecms,aaridgealice,aaridgeatlas,starridge,phenixridge,phobosridge,ppridge,paridge1,paridge2,paridge3,paridge4,paridge5}, has been subject to years of intense theoretical and experimental investigation and can be regarded as a crucial observable in the study of dense QCD.
 The ridge is a 2-particle correlation focused in the azimuthal difference $\Delta \phi$ and elongated in pseudo-rapidity $\eta$.
This correlation seems to be specific to high multiplicity hadronic processes (more central high $\sqrt{s}$ heavy ion,pA,dA \cite{aaridgeatlas,aaridgecms,aaridgealice,phenixridge,starridge,phobosridge}, and very high multiplicity pp collisions \cite{ppridge,paridge1,paridge2,paridge3,paridge4} ), suggesting it to be related to the appearance of a ``medium''.

It is further possible to relate ridge physics to bulk anisotropies particle production \cite{alver,olli,brazil}: The two-particle correlation function, including the ridge, can be successfully decomposed in Fourier components (commonly known as $v_n$) w.r.t. the reaction plane $\Psi_n$
\begin{equation}
\label{vndef}
\frac{dN}{d^3 p} = \frac{dN}{ dp_T^2 d\eta} \left(1+ 2 \sum_{n=1}^\infty v_n(p_T,\eta) \cos\left(n(\phi-\Psi_n) \right)  \right)
\end{equation}
It is trivial to see that a single ridge is mostly generated by $v_{1,3,5}$ components.
A ``double ridge'', observed in AA and pA/dA collisions, is dominated by ``elliptic flow'' $v_2$ and higher even components.    Since elliptic flow was key to the widely-publicized announcement that matter in heavy ion collisions behaves as a low viscosity liquid \cite{sqgp}, a proper understanding of ridge phenomenology is crucial to our understanding of the properties of the matter created in heavy ion collisions.

$v_n$s are thought to reflect the structure of Fourier components of the initial transverse energy density $e(r,\phi,\eta)$ \footnote{I assume boost-invariant longitudinal expansion so the spacetime rapidity and momentum rapidity are the same}.   Thus, $e(r,\phi,\eta)$
\begin{equation}
\label{edef}
e(\vec{x})=e(r,\eta) \left( 1+2\sum_{n=1}^\infty\epsilon_n(r,\eta) \cos\left( n \left( \phi-\Psi_n \right) \right) \right)
\end{equation}
gets somehow converted into momentum anisotropies $v_n$.

Further use of the ridge as a medium probe, however, is somewhat impeded by uncertainty of the mechanism which converts initial state shape anisotropy into a momentum correlation:  It is far from clear if the ridges in $pp,pA,dA,AA$ have the same origin, or a single origin, and how exactly the ridge in each system is generated.

Many models have appeared explaining the ridge.  While experimental data \cite{killmodels} seems to suggest that the ridge is generated by ``soft'' phenomena rather than ``jetty physics'', two broad classes of models remain:  They are distinguished by the ridge's relationship to collective flow.

In the first class of models it is hydrodynamic flow itself that creates the correlation in azimuthal angle \cite{voloshin,moschelli,gavin,hama}.   The local structures (be they hotspots, QCD strings or glasma flux tubes) generate a ``long'' correlation in rapidity which is then azimuthally collimated by transverse flow ``pushing out'' the hotspot.
Thus, the ridge is made by an interplay of transverse flow (the zeroth Fourier component of the flow expansion) and local hotspots, which can be represented by higher harmonics in Eq. \ref{edef}.
The deviation from isotropy, ie the width of the $\Delta \phi$ correlation, should therefore depend directly on the global collective flow of the event and its anisotropies (be they average features, which are there for even coefficients, or event-by-event fluctuations, which can generate odd coefficients).

In the second class of models, the ridge is generated by local partonic dynamics.
As shown in \cite{kovner}, a system  extended in both transverse space and rapidity, approximately boost-invariant, and of high enough partonic density that more than one parton-parton interaction per unit of transverse space per unit of rapidity will naturally yield ridges:
In a spatially extended system, partonic interactions at {\em finite impact parameter} will naturally generate an angular correlation, centered around the impact parameter vector as well as color-coherence regions \cite{kovner2}.   Because of boost-invariance, this angular correlation with be wide in rapidity, since a parton-parton interaction at one rapidity is more likely at a hotspot, and, given that hotspots are elongated in rapidity, a {\em second} parton-parton interaction is more likely.   The model used in \cite{kovner} has been calculated in a generic pQCD model, but qualitatively similar dynamics emerges in CGC-based calculations such as \cite{venu1,venu2,dusling,kovch} as well as string dynamics \cite{paic}.
The scaling behind $e(...)$ is then the local gluon density, given, in \cite{venu1,venu2,dusling,kovch}, by powers of the saturation scale $Q_s$ and the QCD coupling constant $\alpha_s(Q_s)$.
\section{Why the pA ridge is so interesting}
The recent appearance of a ``strong'' $p-A$ and $d-A$ double ridge \cite{paridge1,paridge2,paridge3,paridge4}, comparable to that of heavy ion collisions, has provided us with further puzzles:   Hydrodynamics is generally thought to be a good description of the system when the Knudsen number 
\begin{equation}
\label{knudsen}
Kn = \order{1} \frac{\eta}{s T R} \ll 1
\end{equation}
, i.e. the mean free path is much less than the system size.   
Given the empirical multiplicity scaling with $\sqrt{s}$ \cite{alicemult}, the ideal gas equation of state and the Bjorken model \cite{bjorken} with initial time $\tau_0$, entropy density $s=4(dN/dy)$ and overlap area $S \sim N_{part}^{2/3}$
\begin{eqnarray}
\frac{dN}{dy} = N_{part}  \sqrt{s}^{0.1_{pp,pA},0.15_{AA}} = 4 \tau_0 S s  = 47.5 \frac{8 \pi^2}{45} S \tau_0 T^3
\end{eqnarray}
in terms of the unknown $\kappa = (\eta/(\tau_0 s))$, and assuming the initial radius of a $pp$ collision is set by the strong coupling scale $\lqcd$
\begin{equation} 
Kn_{pp} \propto \kappa (\lqcd)^{-1} s^{0.1} \simeq 0.6 
\end{equation}
assuming $\eta/s = 1/(4\pi)$ and $\tau_0=1fm$.
for larger systems the scaling will be
\begin{equation}
\eqcomma Kn_{pA} \sim N_{part}^{-1/3} K \sim \left(\frac{dN}{dy} \right)^{-1/3} \eqcomma Kn_{AA} \sim N_{part}^{-1/9-1/3} Kn_{pp} s^{-0.02} \sim  \left(\frac{dN}{dy} \right)^{-4/9}
\end{equation}
where the extra energy power in $AA$ is due to the faster dependence of multiplicity per participant in $AA$ collisions \cite{alicemult}.

As can be seen, by comparing ridges at the same $N_{track}$ for different systems, one expects about the same Knudsen number for $pp$ vs $pA$ collisions, but $AA$ will have a lower Knudsen number than an equal multiplicity $pA$ bin:  While the former will be denser but smaller than the other, the greater transverse size of the $AA$ system beats the smaller density in determining $Kn$, and the extra energy dependence will suppress the Knudsen number in $AA$ further.   Therefore, unless one assumes a breakdown in $N_{part}$ scaling \cite{teaney}, binning by multiplicity is {\em not} the same as binning in Knudsen number.

Harmonics should follow a scaling of the type \cite{gtscaling,borghini,lacey}
\begin{equation}
\label{v2kn}
v_n = \order{1} \frac{\ave{c_s}}{\tau_0}  \epsilon_n  \tau_{life}\left(\frac{s_0}{s_f},R \right) \left(1- \order{1} Kn  \right)
\end{equation}
where $\tau_0$ is the hydrodynamic initialization time, $s_f$ is the freeze-out entropy density, $s_0 \simeq (1/S \tau_0)(dN/dy)$ the initial entropy density, $c_s$ the speed of sound and $\tau_{life}$ the lifetime.  

$\tau_{life}$ needs to be determined via a hydrodynamic code but is a non-trivial function of both $s_0$ and the size $R$ even in the ideal limit. It is driven by 
 rarefaction wave dynamics for small ($R \gg T^{-1}$) systems (such as pA,pp) and by expansion dynamics for large ones ($AA$).    
\begin{equation}
\tau_{life}(T,R \gg T^{-1}) \sim \tau_0 \left( \frac{s_0}{s_f} \right)^{1/3} \eqcomma \tau_{life}(T,R \sim T^{-1}) \sim R c_s^{-1}
\end{equation}
Hence, $\tau_{life}$ should further break the Knudsen number scaling.

While,as discussed in \cite{bzdak,mueller}, the initial eccentricity in small systems is too dominated by initial-state uncertainties (the ``transverse shape of the nucleon'') to allow for a quantitative comparison, the scaling patterns of the ridge structure as one goes from pp to pA to AA are puzzling.

If hydrodynamics is to work, Eq. \ref{v2kn} would imply $\eta/(sT) \ll 1$fm, putting the likely value of $\eta/s$ below the ``universal'' $\eta/s =1/(4\pi )$ limit.  
 In this case, however, it is remarkable that
 pp vs AA systems are so different:   
From pA to pp, the Knudsen number should be different by no more than 30-50$\%$, since the transverse size is comparable and the temperatures scale as $T_{pA} \sim N_{part}^{1/3} T_{pp}$.

These considerations arise just from the Knudsen number, and hold independently of the initial geometry.  The latter, for large nuclei, is thought to be dominated by the number of participants.   For smaller systems such as pA and $pp$, however, the largely unexplored sub-nucleon fluctuations are crucial in determining the initial $\epsilon_n$.   
One cannot quantitatively compare $\epsilon_2$ for $AA$ and $pA,pp$ using Eq. \ref{v2kn} since the average geometry in these systems is very different.
Over all events, however, $\ave{\epsilon_3}=0$ by symmetry for {\em all} systems, so event-by-event $\epsilon_3$ is determined only by fluctuations,
\begin{equation}
\ave{\epsilon_3} \sim \sqrt{(\Delta \epsilon_3^{NN} + \Delta \epsilon_3^{sub-N})^2}
\end{equation}
At present nuclear scale $\Delta \epsilon_3^{NN}$ is thought to be well-approximated by a Glauber calculation, while partonic scale $\Delta \epsilon_3^{sub-N}$ is much more model-dependent  \cite{venu1,colemansmith}.    We know, however, that in all models $\left| \Delta \epsilon_3^{NN} \right| \gg \left| \Delta \epsilon_3^{sub-N}\right|$ for $AA$ and  $\left| \Delta \epsilon_3^{NN} \right| \leq \left| \Delta \epsilon_3^{sub-N}\right|$ for pA,pp.    Initial conditions, therefore, give the same hierarchy as the Knudsen number, a ``large'' $AA$ vs a ``small'' $pA,pp$.

Experimental data has a diametrically opposite behavior: The  correlation strength of pA w.r.t. AA collisions are {\em nearly-identical }, especially in $v_3$ where data points at relevant centralities overlap \cite{paridge3}.    For $v_2$, where AA and pA collisions are geometrically different, a $v_2 \sim \epsilon dN/dy$ scaling seems to hold  \cite{paridge1}.    However, when going from AA to $pA$ (LHC) and $dA$ (RHIC), the Knudsen number certainly rises by $\sim \order{2-4}$ or so if transverse size is counted as $R$.    

At present, only one experiment \cite{ppridge} reported seeing a ridge in $pp$ collisions, and only in an experiment-specific sample of very high multiplicity events.    
This ridge is of markedly smaller amplitude w.r.t AA, and might not exhibit the accompanying double ridge (which for pA was seen in all centrality bins and can be used to estimate $v_2$ and $v_3$).    While no harmonic decomposition was attempted to date, $v_2$ and $v_3$ in pp collisions could be compatible with zero, albeit with a large error bar.

This large systematic uncertainty is primarily due to the large background, in $pp$ collisions, of off-center of mass energy (different Bjorken $x$) parton-parton collisions producing a strong away-side ``fake ridge'' \cite{jurgen}.
Unlike in pA collisions, the admixture of such events is highly correlated with multiplicity, making their subtraction difficult.

Hence, a centrality-based quantitative comparison is to date problematic. 
Even a qualitative comparison between pp and pA ridges \cite{paridge3} however, shows it is difficult to see how the Knudsen and density scaling described in the previous paragraph can be made compatible with experimental data.

I should note that, as pointed out earlier, this is a more general issue than the pA ridge.   It seems (see the references in \cite{gtscaling,gtscaling2} that when different energies, system sizes and rapidities are put together, $v_2$ is described by the following empirical formula
\begin{equation}
\label{v2scaling}
v_2(p_T) \simeq \epsilon_2 F(p_T) \eqcomma \ave{v_2} \simeq \frac{1}{S}\frac{dN}{dy} = \int dp_T F(p_T) f\left(\frac{p_T}{\ave{p_T}} \right)
\end{equation} 
where $F(p_T)$ and $f(p_T/\ave{p_T})$ are universal functions (with very weak to no dependence on $\sqrt{s},A,N_{part},y$ and so on) and the residual dependence on $(1/S)(dN/dy)$ is through $\ave{p_T}$ rather than $v_2(p_T)$.   

While a detailed study of this scaling within hydrodynamics is yet to be performed, the compatibility of such a system with a fluid whose speed of sound $c_s$ and $\eta/s$ depend non-trivially on temperature is not so obvious.  Expanding the Cooper-Frye formula \cite{cf} (discussed later in Eq. \ref{cf}) in the 2nd Fourier Harmonic of the flow and freeze-out hypersurface 
\begin{equation}
\label{v2pt}
v_2(p_T) \simeq \int d\phi \cos^2 (2\phi) \left[ \underbrace{e^{ - \frac{\gamma\left(E-p_T v_T \right)}{T}}}_{v_T \simeq c_s \tau_{life}(...)} \left(1 - p_T \underbrace{\Delta
\frac{dt}{dr}}_{\sim \epsilon_2}  + \underbrace{\frac{ \delta v_T p_T }{T}}_{\sim \epsilon_2} + \order{\epsilon^2,Kn} \right) \right]
\end{equation}
which, when integrated, yields Eq. \ref{v2kn} ($\tau_{life}$, defined after Eq. \ref{v2kn}, also determines $v_T$). 
   Hence, size, density and $p_T$ dependence are {\em not} expected to factorize they do in Eq. \ref{v2scaling}, but maintain a non-trivial dependence in which density and size (and hence geometry, $N_{part}$, system size and rapidity) mix in a non-trivial way.
\begin{equation}
\left( \begin{array}{c}
v_2(p_T) \\
\ave{p_T}\\
\frac{1}{S}\frac{dN}{dy}
\end{array} \right)= \tau_{life}\left( \frac{s_0}{s_f},R \right) 
 \times
\left( \begin{array}{ccc}
\order{1} &- \order{1} & 0 \\
  \order{1} & \order{1} & \order{1} \\
  \order{1} & \order{1} & \order{1} 
\end{array}
 \right) \times \left( \begin{array}{c}
\epsilon_2 \\
Kn\\
\order{1}
\end{array} \right) 
\label{matrix}
  \end{equation}
If all $\order{1}$ parameters are non-negligible and uncorrelated, no projection of the matrix in Eq. \ref{matrix} to a lower dimension is possible.   The only way to reduce the dimensions of Eq. \ref{matrix} is to assume a negligible Knudsen number.   But, as discussed before, this predicts similar correlations between p-p and pA and does not eliminate scaling violations associated with $\tau_{life}$.
Hence, the scaling difficulty of the pA ridge is an ``in your face'' (due to pA collisions's small size) illustration of the larger difficoulty of hydrodynamics to describe the {\em scaling} of harmonic flow with energy,system size and rapidity \cite{gtscaling,gtscaling2}.  

In contrast, initial state effects depend only on the geometry {\em and only one intensive parameter} (such as the ``saturation scale'' $Q_s$, or more generally the transverse gluon density \cite{kovner}). 
Thus, such models might be more amenable to simple scaling \cite{larryid,larry1}. 
The long-established difference between in-vacuum and in-medium partonic wavefunctions \cite{emc} also makes the similarity of pA and AA w.r.t. pp more natural in such an initial state model.
   The systematic uncertainties over the initial geometry in small systems, however, make disentangling such initial state effects from collectivity remnants just from particle harmonics in a model-independent way  non-trivial.

From a scaling point of view, therefore, hydrodynamics looks disfavored but the entanglement of geometric and ``intensive'' quantities might be too high for a ``clean'' to be made by scaling arguments {\em alone}.
This puts the onus on exploiting the main difference between the two models of ridge production outlined in the previous section - in one case azimuthal collimation is done by ``global'' collective flow, in the other by ``local'' parton-parton interactions, to devise a qualitative experimental observable which can clarify the situation.   In the next section I will argue that PID two-particle correlations is a good candidate for such an analysis.
\section{Why PID correlations are important}
In $AA$ collisions, collective flow has historically been ascertained by looking at the difference between $\pi,K,p$ spectra.
When an expanding hydrodynamic fluid emits particles, energy-momentum and entropy conservation constrain the particle distribution to be of the Cooper-Frye form
\begin{equation}
\label{cf}
E \frac{dN}{d^3 p} = \int S(\vec{x},\vec{p})d^3 x
\end{equation}
where the Emission function is a ``boosted'' thermal exponential \cite{cf}.
\begin{equation}
\label{emission}
S(x,p)= \frac{d \Sigma_\mu}{d^3 x} p^\mu \exp\left [-\frac{p_\mu u^\mu(\vec{x})}{T(\vec{x})} \right] + \order{Kn \times \left[ \frac{p_T}{T} \right]^{3/2-2}}
\end{equation}
the form of the viscous correction is currently controversial \cite{cfv1,cfv2,cfv3}, but it is subleading in Knudsen number and important at $p_T \gg T$.    The $Kn \rightarrow 0$ limit is the usual boosted exponential
\begin{equation}
\label{emission2}
S(x,p)   \sim  \ave{\exp \left[ - \frac{\gamma}{T(r)}\left(E-p_T v_T(r,\phi) \right)  \right]}_r
\end{equation}
where $T$ is the temperature and $v_T$ is the transverse flow at freeze-out (the radial component of the flow 4-velocity $u^\mu$) and $\Sigma_\mu$ is the 3D surface specifying the spacetimes of particle emission.
$T(x),u_\mu(x),d \Sigma_\mu$ have to be calculated within a hydrodynamic code.
In this work, I quote the results in \cite{us} (a similar analysis was done in \cite{them}) calculated with the code in \cite{bozekhydro}, which was also used successfully to describe small systems \cite{bozek} in the past.

A qualitative signature of Eq. \ref{cf} is a mass-scaling of spectra and flow correlations.  For a particle of mass $m$, the exponent at mid-rapidity (no longitudinal momentum) becomes
\begin{equation}
\alpha= \frac{\sqrt{p_T^2+m^2} - v_T p_T}{p_T} \simeq \left\{  
\begin{array}{cc}
 (1-v_T) + \frac{m^2}{p_T^2} & p_T \gg m\\
 \left(\frac{p_T}{2m}  -v_T \right) +\frac{m}{p_T} & p_T \ll m 
\end{array}
 \right.
\end{equation}
Consequently, each particle still maintains an exponential shape, but its ``effective temperature'' $T'$ is related to the true freeze-out temperature $T$ by
$T' \simeq T (\gamma/\alpha(m))$
the $m$-dependence is steepening in both the massive and the massless particle limit, with the spectrum becoming flatter as $m$ increases (naturally, since more momentum is transmitted to the particle by flow and less by thermal motion).

Thus, mass ordering arises naturally if angular correlations are generated by flow, as assumed in the first class of scenarios discussed in the introduction.
If, however, the second class is physically correct, and the angular correlations are generated by initial state high density effects, mass ordering can also arise from final state fragmentation effects.
As noted in \cite{dumitru}, a non-negligible mass-ordering is generated within PYTHIA with no need for transverse flow, by the fact that more massive particles are typically produced by more ``transverse'' strings.   For small (p-p) systems, mass ordering using this effect is {\em greater} than the mass ordering from a hypothetical hydrodynamic phase.   

While \cite{dumitru} focused on $q \overline{q}$ strings, while the processes considered in \cite{dusling} are purely gluonic,
such effects should also be present in fragmentation functions: given a distribution of gluons $f(p)$ fragmenting as in elementary collisions, the final hadron distribution will be given by
\begin{equation}
\label{fragment}
\frac{dN_i}{d^3 p} = \int f\left( p_g,\phi_g \right) D_{g \rightarrow i}\left(\frac{p}{p_g},p_g^2  \right) d^3 p_g
\end{equation}
different $D_{g \rightarrow i}$ will generate different hadron spectra.  It is therefore not surprising that models such as HIJING reproduce qualitatively, although not quantitatively \cite{us} the $\ave{p_T}$ scaling in pA collisions.
Since our understanding of gluon fragmentation is far from precise, therefore, the mass hierarchy, {\em by itself}, cannot be taken as evidence of hydrodynamic flow, although an extension of $\ave{p_T}$ analysis for particles of similar masses but different partonic wavefunctions, such as the proton and the $\phi$ meson, might help in clarifying whether the scaling variable in $\ave{p_T}$ is mass or partonic structure.

2-particle correlations are however qualitatively different from 1-particle correlations in this respect.
If the ridge is generated by transverse flow focusing hotspots, its strength should be correlated, and its width anti-correlated, with the mass of the particle, because of the greater sensitivity to flow of higher-mass particles.   This leads to higher, slimmer correlations for more massive particles.   Mathematically, \cite{cf} accounts for this effect by 
modeling 2-particle angular correlations through the local flow field.  Up to an emission volume $\mathcal{V}$, and ignoring quantum (HBT) effects
\begin{equation}
\label{cf2part}
\frac{d_{ij} N}{dp_1 dp_2 d\phi_1 d\phi_2} \propto \mathcal{V} \int S(\vec{x},\vec{p_1}) S(\vec{x},\vec{p_2}) d^3 x 
\end{equation}
with $S(...)$ specified in Eq. \ref{emission} and the mixed event background given by the square of Eq. \ref{cf}. 
 It is clear from the form of Eq. \ref{emission} that correlations will get an additional boost for higher-mass particles, since the latter are more sensitive to flow.    This is why hydrodynamics predicts consistently higher $v_2$ form more massive particles \cite{huovinen}.

As an illustration, \cite{us} calculated the 2-particle distribution function, binned by particle species, with the model \cite{bozek}.
The result is shown in Fig. \ref{figcorr} for
\begin{equation}
\label{cdef}
C(\Delta \phi) =  \frac{1}{N_{tot}} \int\left( \frac{d_{ij}^2 N}{dp_1 dp_2 d\phi_1 d\phi_2} - \frac{dN_i}{ dp_1 d\phi_1}\frac{dN_j}{dp_2 d\phi_2} \right) dp_{1,2} d\phi_{1,2} \delta\left( \phi_1 - \phi_2 - \Delta \phi \right)
\end{equation}
where $N_{tot}$ is the overall number of events, and indeed follows expectations.   Experimental data seems to behave in a similar way \cite{us,paridge5}.

In contrast, in initial-state based models \cite{kovner,dusling,kovch} the effective theory of high-density QCD will give a 2-parton correlation.    In the absence of subsequent evolution, however, fragmentation will happen ``as in vacuum'', or independently for each gluon.    gluon-gluon correlations will be suppressed by factors of $\alpha_s$, assumed as small in such approaches.
The hadron-hadron ($i-j$) correlation function will therefore be of this form
\begin{equation}
\label{fragment2part}
\frac{dN_{ij}}{d^3p_1 d^3p_2} \propto  \int f_2\left( p_{g1},p_{g2},\phi_{g1}-\phi_{g2} \right) D_{g \rightarrow i}\left(\frac{p_1}{p_{g1}},p_{g1}^2  \right) D_{g \rightarrow j}\left(\frac{p_2}{p_{g2}},p_{g2}^2  \right) d^3 p_{g1} d^3p_{g2} 
\end{equation}
where $f_2\left( p_{g1},p_{g2},\phi_{g1}-\phi_{g2} \right)$ is a function incorporating the predictions from dense parton dynamics for the ridge-like structure.   The mixed event background is given by the square of the distribution in Eq. \ref{fragment}.

Even qualitatively, Eq. \ref{fragment2part} is a very different equation from Eq. \ref{cf2part}, since fragmentation is parton-specific and independent of and partonic correlations.    Thus, the fragmentation function $D_{i\rightarrow j}(...)$ is independent of rapidity and angle.   Hence, for the azimuthal correlation function all dependence on $D(....)$ factorizes into a function of $p$ only.   Provided normalization is done by a species and $p-$bin specific factor so one point on the curve matches for all species, any residual fragmentation effects should factor out.    This should {\em not} be the case if particle production happens via Cooper-Frye freezeout, Eqn \ref{cf2part}.

\begin{figure}[t]
\epsfig{width=17cm,figure=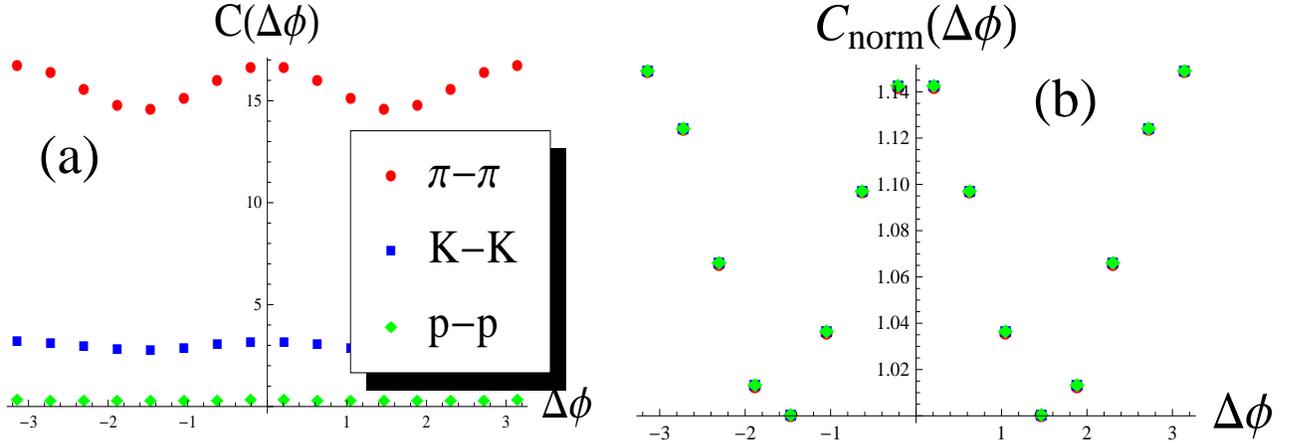}
\caption{\label{figmar} (color online)
2-particle correlation function, binned by particle species with different normalizations (panel (a)) and normalized to a common minimum determined by ZYAM for every particle (panel (b)).  The correlation function was calculated using Eq. \ref{fragment2part} and empirical distribution Eq. \ref{empirical}}
\end{figure}

To test the phenomenological consequences of this, I describe 
$ f_2\left( p_{g1},\phi_{g1},\eta_{g1},p_{g2},\phi_{g2},\eta_{g2} \right)$ by an empirical function in terms of $\Delta \phi$ and $\Delta \eta$ (the narrow peak in $\Delta \phi,\Delta \eta$ due to jet fragmentation is disregarded). 
\begin{equation}
 f_2\left( p_{1},\phi_{1},\eta_1,p_{2},\phi_{2},\eta_2 \right) = \left( 1+ A \cos\left( 2 \left( \phi_1 - \phi_2 \right) \right) \right)
 \times \tanh\left( \frac{p_1}{B} \right) \tanh\left( \frac{p_2}{B} \right) 
\label{empirical}
\end{equation}
e $A,B$ are adjustable parameters
  and proceed to convolute it with the CTEQ \cite{cteq} gluon fragmentation functions according to Eq. \ref{fragment2part}.
\begin{figure}[t]
\epsfig{width=15cm,figure=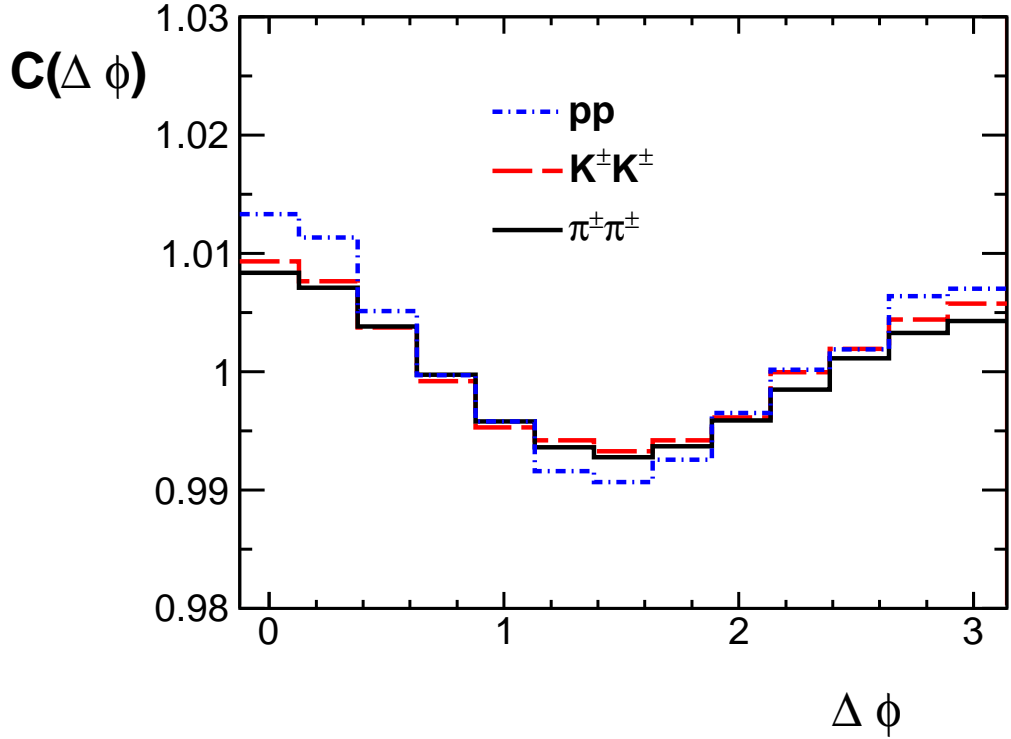}
\caption{(color online) \label{figcorr}  2-particle correlation function, binned by particle species, calculated in a hydrodynamic model using Eq. \ref{cf2part}.  
Plot originally published in \cite{us}.
}
\label{figcorr}
\end{figure}
The result is also shown in Fig. \ref{figmar}, for both a normalization of the type $C(\Delta \phi)$ and assuming Zero Yield At Minumum (ZYAM) separately for all particle species, $C_{norm}(\Delta \phi)$ ($C_{norm}=1$ at minumum by definition).  As can be seen, once an overall normalization factor is common for all particles, the correlation function shape is independent of the particle species.  
This is clearly not the case for the plot in Fig. \ref{figcorr}.

  Changing the form of the empirical function Eq. \ref{empirical} without the hadronization implied in Eq. \ref{fragment2part} will not alter these basic conclusions, although of course the shape of the angular correlation and its normalization will vary.

Physically, the species-independence of the shape of the 2-particle correlation function is a trivial consequence of the form of Eq. \ref{fragment2part}: since fragmentation into a particular species are determined independently of the correlation, the fragmentation function factorizes into ``the average energy of the gluon that fragments into a hadron of that momentum'',which retains no memory of the shape of the angular distribution.  

By how much can this feature be altered by altering fragmentation dynamics?
Collinear fragmentation assumed in Eq. \ref{fragment2part} is usually appropriate for $p_T \gg \lqcd$.   For lower momenta, fragmentation is certainly not collinear.  However, if the typical parton momentum $\sim Q_s > \lqcd$ fragmentation should remain parton specific since correlations between oppositely moving partons are higher order in $\alpha_s(Q_s)$.

One way to incorporate this is to introduce species-specific deviations from collinear fragmentation, i.e.,
$D_{g \rightarrow i}\left(\frac{p_i}{p_{g}},p_{g}^2  \right) \Rightarrow D_{g \rightarrow i}\left(\frac{p_i}{p_{g}},p_{g}^2,\phi_g - \phi_i  \right)$.  The $\phi_g-\phi_i$ dependence might introduce the observed species-specific pattern in the 2-particle correlation function if (as is reasonable) fragmentation into more massive particles gives a larger ``side-kick'' to the fragmented hadron.

Assuming a normalization such that $$\int_0^{2\pi}\int_0^1 D(z,\phi_i-\phi_g)dz d\phi_i=1$$
and a gluon distribution correlated by $v_2$ only
\begin{equation}
\label{vncorr}
C\left( \phi_{1g}-\phi_{2g} \right)   \propto 1+ \sum_n  v_{gn}^2 \cos\left( n \left(\phi_{g1} - \phi_{g2} - 2\Psi_n \right) \right)
\end{equation}
one can see that, when one integrates $d^3 p_{g2}$ out of Eq. \ref{fragment2part}
one gets 
\begin{equation}
\label{vn1part}
\frac{d N_{i} }{d p_{i} d\phi_{i}} \propto \int d\phi_g dp_{Tg} \left(  1 + 2\sum_n v_{ng}(p_{Tg}) \cos(n \left( \phi_g -\Psi \right) \right) D_{g \rightarrow i} \left( \frac{p_{Ti}}{p_{Tg}},\phi_i-\phi_g   \right)
\end{equation}
Note that no correlation exists between $\phi_g-\Psi$ and $\phi_i-\phi_g$, integrating over one of these angles for a large ensemble of both events and fragmentations will average the other to zero.
Hence, Fourier-decomposing Eq. \ref{vn1part} one can rotate $\Psi$ or $\phi_g$  away into an irrelevant phase. Using the convolution theorem, one will get
\begin{equation}
\label{dtilde}
v_n^{hadron}= \int dp_{Tg} v_{ng}(p_{Tg}) \frac{dN_{g}}{dp_{Tg}} \int dp_{Ti} \ave{\tilde{D}\left( \frac{p_{Ti}}{p_{Tg}} \right)}    \simeq \ave{v_{ng}}\ave{\ave{ \tilde{D}_{g\rightarrow i}}}
\end{equation}
where $\tilde{D}$, the Fourier-transform of the product of fragmentation function, and {\em must be independent of geometry and $n$-coefficient} (In the collinear limit $\ave{\tilde{D}_{g \rightarrow i}} \rightarrow 1$, in case fragmentation smearing dominates  $\ave{\tilde{D}_{g \rightarrow i}} \rightarrow 0$) .     
In contrast, $v_n^{gluon}$ must depend only on geometry and not of particle species.
Ratios of $v_n$s for different system sizes and particle species, a la \cite{larryid},therefore, could be used to see if fragmentation of correlated gluons occurs independently for each gluon or is correlated.   If relation Eq.\ref{dtilde} holds, for two systems $A,A'$ (be they $pp,pA,dA$ or a particular centrality of$AA$) at the same $\sqrt{s}$ and $p_T$
\begin{equation}
 \left. \frac{v_2^\pi}{v_3^\pi} \right|_{A} \simeq \left. \frac{v_2^p}{v_3^p} \right|_{A}\eqcomma  \frac{v_n^\pi|_A}{v_n^\pi|_{A'}} \simeq  \frac{v_n^p|_{A}}{v_n^p|_{A'}} \eqcomma \left. \frac{v_3^p}{v_3^\pi} \right|_{A} \simeq \left. \frac{v_3^p}{v_3^\pi} \right|_{A'} 
\label{tests}
\end{equation}
where the last relation is motivated by the nearly identical $v_3$ in pA/dA vs AA collisions \cite{paridge3}.
No such relations are apparent in the hydrodynamic model. In fact the last two ratios in Eq. \ref{tests} should exhibit a systematic upward trend for larger systems, as transverse flow is larger in such systems (As the system size increases, $v_n^p$ grows faster than $v_n^\pi$).
This is something experiment can readily check.

If such scalings are not apparent, initial-state correlations might still work 
 if the multi-gluon ladders commonly used to justify the initial-state $k_T$ factorization approach \cite{kt1,kt2}
also influence fragmentation, leading to some fragmentation function sensitive to multi-gluon correlations
\begin{equation}
 D_{g1 \rightarrow i}\left(\frac{p_1}{p_{g1}},p_{g1}^2,\phi_i - \phi_{g1}  \right) D_{g2 \rightarrow j}\left(\frac{p_2}{p_{g2}},p_{g2}^2,\phi_j-\phi_{g2}  \right) \Rightarrow 
\label{corrfrag}
\end{equation}
\[\ \Rightarrow D_{g_{1,2} \rightarrow ij}\left(\frac{p_1}{p_{g1}},p_{g1}^2,\frac{p_2}{p_{g2}},p_{g2}^2,\phi_i-\phi_{g1},\phi_j-\phi_{g2}  \right) 
\]
Such a fragmentation would be make the experimental distinction between an initial-state model and a hydrodynamic model much more involved.

The good description, by the IP-Glasma flux-tube model with parton-hadron duality, of the Negative-Binomial multiplicity distributions \cite{Gelis:2009wh} places some bounds of the importance of such effects within an initial state-based model.
Nevertheless, these ideas have been explored in the literature before \cite{junction1,junction2}, although they have not been explored quantitatively in the context of 2-particle correlations.   
If multi-gluon correlations persist until and including fragmentation, however, it is difficult to see the qualitative difference between this system and a dense ensemble of sequentially interacting gluons, which should universally approach hydrodynamic evolution \cite{bamps}.   In this limit, ``high density gluons'' and ``hydrodynamics'' become indistinguishable.
\section{ Discussion and conclusions}
Further tests of the observables described here can be experimentally performed via 2-particle correlations of the $\phi$-meson, which has the approximately same mass as the baryon but a very different partonic structure: A similar pattern to the proton would point to mass scaling, typical of hydrodynamics and Equation \ref{cf2part}.  A very different pattern might indicate non-hydrodynamic fragmentation, possibly correlated fragmentation of the type of Eq. \ref{corrfrag} if the tests outlined in Eq. \ref{tests} are not passed. 

Along the same direction, 2-particle correlations with identified {\em heavy quark mesons} in pA and dA collisions might be instrumental in assessing the relative importance of flow w.r.t. initial state correlations:  While correlations due to high initial gluon density should be equally strong for light and heavy quarks according to formulae similar to Eq \ref{fragment2part} (updated with the heavy quark creation diagrams \cite{tuchin}), in a hydrodynamic system the ``effective Knudsen number'' for a heavier particle is $Kn_{M} \sim \frac{M}{T} Kn$\cite{moorem}, so a hydrodynamic correlation will be parametrically weaker.  

It will also be illuminating to see whether $v_n(p_T)$ in pA go to the same $p_T \simeq 50$ GeV as in AA \cite{cmshighpt}.  For pA the most commonly used tomographic variable, $R_{pA}(p_T) \simeq 1$ from $p_T \simeq 3-4$ GeV onward \cite{alicerpa}.   This would naively suggests the ``medium'' in pA is transparent to particles at 3-4 GeV. Yet $v_2(p_T)$ in pA was measured to be non-negligible for these momenta \cite{paridge3,paridge4,paridge5}.    This suggests that the influence of ``the medium'' on 1 and 2 particle correlations is dramatically different.   This observation, as well as giving us an explanation for the inability of most jet energy loss models to describe both $R_{AA}$ and $v_2$ together (unless a non-trivial opacity evolution with density is assumed \cite{liao,betz}), raises the question of how far in $p_T$ does $v_n(p_T)$ reach in ``small systems'' such as pA and dA collisions.   

To answer these questions, one must determine if $p_T=3-4$ GeV in the ``soft'' regime determined by hydrodynamics or the ``hard'' regime determined by fragmentation.
Knudsen number scaling suggests, as shown in the appendix, that for a ``fluid'' with low absolute opacity, after a 
\begin{equation}
\label{ptkn}
p_T \gg \left(T^3 \epsilon_n R/Kn \right)^{1/2-2/3}
\end{equation}
, one expects hydrodynamic $v_n(p_T)$ to be $\simeq 0$.   
Tomographic $v_n$ for these $p_T$s is ruled out by $R_{pA} \simeq 1$.

A sizable $v_2$ for $pA/dA$ collisions at $p_T \gg 10$ GeV might suggest that either $\eta/s$ is {\em really} low (but then again, what about pp collisions?), or initial state is somehow impacting 2-particle correlations even at high momenta.

In conclusion, I have discussed the possible physical origin of the observed ``ridge-like'' 2-particle correlations observed in pA and AA collisions.
I cannot draw any firm conclusions:  On the one hand, the scaling of the observed ridge as the system size increases from pp to pA/dA and AA is difficult to see in a hydrodynamic model.    On the other, the successful fit by hydrodynamic calculations \cite{us,them} of the PID ridge, together with the failure, even on a qualitative level, of existing in-vacuum fragmentation ansatze to reproduce the pattern observed in experiment, suggests that the ridge is very much a {\em final state} phenomenon:
The mass ordering, provided fragmentation happens through usual vacuum fragmentation, is non-trivial to reproduce when the 2-gluon correlation function is tuned to reproduce charged dihadron correlations.
I eagerly hope that further tests for this mass ordering, involving $\phi$ mesons, heavy quarks and high $p_T$ charged particles, can clarify these issues.
\section{Appendix: The ``effective Knudsen number'' at higher $p_T$}
Combining the standard formula $\frac{\eta}{s} \sim T l_{mfp} \sim \frac{1}{\ave{\sigma(p_T) T^2}}$ with the dimensional estimate $\sigma(p_T\gg T)/\ave{\sigma} \sim T^2/p_T^2$ and the Knudsen number defined in Eq. \ref{knudsen} I get that, for $p_T \sim \sqrt{T^3 R/Kn}$ the number of ``hard'' scatterings is $\sim 1$.
For a Fourier component in the scattering difference in azimuthal angles to vanish, a it is enough that the azimuthal difference in number of scatterings be small.  For this one requires $p_T \sim \sqrt{T^3 \epsilon_n R/Kn}$ for the $n$th component.  A radiative dominated freezeout will bring the square root to a 3/2 root \cite{cfv3}.
  ``Tomographic'' soft scatterings and radiative corrections could remain, as they do not contribute to the transport properties of the bulk per se. However, $R_{pA}\simeq 1$ after a $p_T =3-4$ GeV seems to place strict limits on their significance.

G.T. acknowledges the financial
support received from the Helmholtz International Center for FAIR within the framework of the LOEWE program (Landesoffensive zur Entwicklung
Wissenschaftlich-\"Okonomischer Exzellenz) launched by the State of Hesse.
GT also acknowledges support from DOE under Grant No. DE-FG02-93ER40764. 
I would like to thank 
Wojciech Broniowski, Piotr Bozek, Miklos Gyulassy, Mike Lisa and Jurgen Schukraft for discussions.

\end{document}